\documentclass[lettersize,journal]{IEEEtran}
\usepackage{amsmath,amsfonts}
\usepackage{algorithmic}
\usepackage{algorithm}
\usepackage{array}
\usepackage[caption=false,font=normalsize,labelfont=sf,textfont=sf]{subfig}
\usepackage{textcomp}
\usepackage{stfloats}
\usepackage{url}
\usepackage{verbatim}
\usepackage{graphicx}
\usepackage{cite}
\usepackage{multirow}
\usepackage{booktabs}
\usepackage{float}
\usepackage{color}

\newcommand{\etal}{\textit{et al}.}
\newcommand{\ie}{\textit{i}.\textit{e}.}
\newcommand{\eg}{\textit{e}.\textit{g}.}

\begin{document}

\title{Unleashing Correlation and Continuity for Hyperspectral Reconstruction from RGB Images}

\author{Fuxiang Feng, Runmin Cong,~\IEEEmembership{Senior Member,~IEEE,} Shoushui Wei, 
Yipeng Zhang,
Jun Li,~\IEEEmembership{Fellow,~IEEE,}\\ Sam Kwong,~\IEEEmembership{Fellow,~IEEE,} and Wei Zhang,~\IEEEmembership{Senior Member,~IEEE}

\thanks{Fuxiang Feng, Runmin Cong, Shoushui Wei and Wei Zhang are with the School of Control Science and Engineering, Shandong University, Jinan 250061, China, and also with the Key Laboratory of Machine Intelligence and System Control, Ministry of Education, Jinan 250061, China (e-mail: fuxiangfeng@mail.sdu.edu.cn; rmcong@sdu.edu.cn; sswei@sdu.edu.cn; davidzhang@sdu.edu.cn).}
\thanks{Yipeng Zhang is with the University of California, Los Angeles, America
 (e-mail: zyp5511@g.ucla.edu).}
\thanks{Jun Li is with the Hubei Key Laboratory of Intelligent Geo-Information Processing and the School of Computer Science, China University of Geosciences, Wuhan 430078, China (e-mail: lijuncug@cug.edu.cn).}
\thanks{Sam Kwong is with the Lingnan University, Hong Kong SAR, China (e-mail: samkwong@ln.edu.hk).} 
\thanks{This work has been submitted to the IEEE for possible publication.
Copyright may be transferred without notice, after which this version may no longer be accessible.}
} 

\markboth{IEEE Transactions on XXX}%
{Shell \MakeLowercase{\textit{\etal}}: A Sample Article Using IEEEtran.cls for IEEE Journals}

\maketitle

\IEEEpubid{\begin{minipage}{\columnwidth}\raggedright
    Copyright \copyright 2025 IEEE.
\end{minipage}}
\IEEEpubidadjcol 

\begin{abstract}
Reconstructing Hyperspectral Images (HSI) from RGB images can yield high spatial resolution HSI at a lower cost, demonstrating significant application potential. 
This paper reveals that local correlation and global continuity of the spectral characteristics are crucial for HSI reconstruction tasks. 
Therefore, we fully explore these inter-spectral relationships and propose a Correlation and Continuity Network (CCNet) for HSI reconstruction from RGB images.
For the correlation of local spectrum, we introduce the Group-wise Spectral Correlation Modeling (GrSCM) module, which efficiently establishes spectral band similarity within a localized range.
For the continuity of global spectrum, we design the Neighborhood-wise Spectral Continuity Modeling (NeSCM) module, which employs memory units to recursively model the progressive variation characteristics at the global level.
In order to explore the inherent complementarity of these two modules, we design the Patch-wise Adaptive Fusion (PAF) module to efficiently integrate global continuity features into the spectral features in a patch-wise adaptive manner.
These innovations enhance the quality of reconstructed HSI.
We perform comprehensive comparison and ablation experiments on the mainstream datasets NTIRE2022 and NTIRE2020 for the spectral reconstruction task. 
Compared to the current advanced spectral reconstruction algorithms, our designed algorithm achieves State-Of-The-Art (SOTA) performance.

\begin{IEEEkeywords}
hyperspectral reconstruction, spectral correlation, spectral continuity, attention, convolution neural network.
\end{IEEEkeywords}

\end{abstract}

\section{Introduction}\label{sec1}

\IEEEPARstart{H}{yperspectral} Images (HSI), celebrated for their rich spectral characteristics, capture significantly more semantic information compared to RGB images.
This advantage has made HSIs highly valuable in diverse fields, such as remote sensing analysis, environmental monitoring, and mineral identification \cite{luo2018feature, su2022hyperspectral}.
Moreover, their distinctive spectral properties have positioned HSIs as a focal point in computer vision, driving advancements in related technologies, including image classification \cite{su2022nsckl, zhang2020symmetric}, target detection \cite{dong2024deep, li2022target}, semantic segmentation \cite{lu2013spectral, fang2023hyperspectral}, among others.

Despite their promise, the predominant method of acquiring HSIs involves capturing them directly from the environment using hyperspectral cameras, a process accompanied by several limitations. 
First, due to the limitations of spectral imaging mechanisms, the spatial resolution of the raw HSI captured by hyperspectral cameras is relatively low.
Additionally, to ensure sufficient energy capture for each spectral band, the sensor requires prolonged exposure times, leading to slow imaging speeds \cite{james2007spectrograph}. 
Furthermore, the high manufacturing costs of professional hyperspectral cameras pose significant barriers to their widespread practical application.
These limitations hinder the broader adoption of hyperspectral image processing technologies in real-world scenarios. 
Therefore, proposing a high-resolution, high-timeliness and low-cost hyperspectral imaging approach is highly meaningful.

Considering the low-cost and efficient imaging capabilities of RGB images, many researchers have started exploring methods to transform RGB images into HSI. 
As a result, the task of HSI reconstruction from RGB images has emerged. 
But this process involves converting from three-dimensional space to higher-dimensional space, which is inherently ill-posed.
Traditional algorithms for reconstructing HSI primarily rely on prior knowledge to establish mapping relationships from RGB to HSI \cite{liu2015improving, arad2016sparse}. 
However, these methods often have limited feature representation capabilities, posing challenges for their practical application and generalization. 
With the development of deep learning, Convolutional Neural Network (CNN) algorithms have gained increasing attention in HSI reconstruction tasks and achieved considerable performance \cite{alvarez2017adversarial, koundinya20182d, shi2018hscnn+, zhao2020hierarchical,li2020hybrid, zheng2021spectral, li2022hasic, li2022drcr, zhao2023hsgan, mei2023lightweight}.
In order to model the correlation between adjacent spectral bands, some researchers have combined 1D CNN with channel attention \cite{zheng2021spectral, mei2023lightweight} or employed 3D CNN \cite{koundinya20182d, li2020hybrid} to reconstruct spectral features.
Recently, due to the characteristics of Transformer in modeling long-range dependencies, the Transformer-based network has become a more advantageous and popular solution.
For example, Cai \etal \cite{cai2022mst++} were pioneers in treating the band dimension of spectral features as a sequence dimension, and employed Multi-Head Attention (MHA) module to model the similarity correlations of spectral features between each band and all other bands.
Compared to algorithms that solely utilize CNN, these algorithms can explore more complex inter-spectral feature dependencies.

\begin{figure*}[!]
        \centering
	\includegraphics[width=0.90\textwidth]{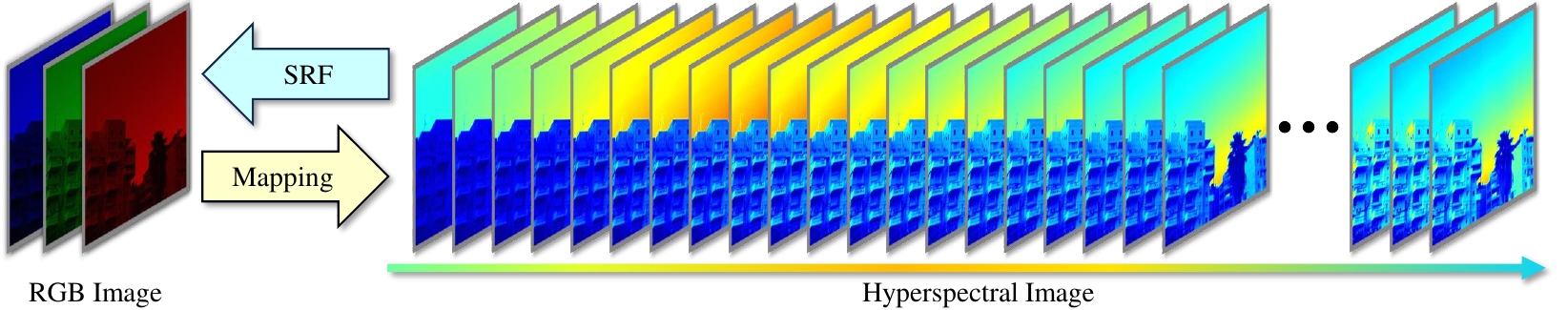}
	\caption{{The process of spectral reconstruction from RGB image. The RGB image has three channels: R (red), G (green), and B (blue), which are perceivable by the human eye. The hyperspectral image composed of multiple bands, and the images from left to right represent the visualization of response values from low to high across different wavelength.
    }}
	\label{process}
\end{figure*}

In summary, previous improvements have primarily focused on constructing similarity associations between spectral features.
However, their analysis of inter-spectral characteristics remains incomplete.
It is well known that the HSI process involves equidistant sampling within a continuous spectral band range.
Therefore the collected spectral response data, as depicted in Fig. \ref{process}, exhibit two key characteristics: local correlation and global continuity.
Firstly, the inter-spectral relationship shows the characteristics of strong local correlation.
The spectral similarity between adjacent bands is higher, while it decreases for bands with larger intervals.
In previous work, CNN-based approaches have not effectively constructed inter-spectral correlations. 
Although MHA module can achieve high-order correlation modeling, it does not emphasize the construction of correlations between local spectral bands. 
Secondly, from a global perspective, the spectral response value of HSI at each spatial position changes progressively along the global band dimension, and we refer to this phenomenon as the global continuity characteristic. 
However, previous methods have not modeled such progressive changing inter-spectral properties in the reconstruction process.
To this end, we fully exploit the inter-spectral relationship characteristics and propose a Correlation and Continuity Network (CCNet) for HSI reconstruction.

To address the aforementioned issues, it is essential to consider how to efficiently model the inter-spectral relationships, including correlations at local level and continuity at global level.
On the one hand, in order to achieve correlation modeling of local spectrum in an efficient manner, 
we abandon the one-to-all modeling mode of the MHA mechanism and design a group-wise local inter-spectral modeling strategy. 
Specifically, we propose the Group-wise Spectral Correlation Modeling (GrSCM) module to achieve group-wise attention operations, which uses image-level feature fusion method within local regions to model inter-spectral correlation.
The spectral segments in the local adjacent space are more fully associated, which further emphasizes the strong local correlation characteristics between spectral features.
Additionally, by refining the attentional scope, we reduce the number of attention weights that need calculation and alleviate the network's optimization burden. 
On the other hand, we consider the spectral features as a time series along the band dimension and intuitively adopt a recursive approach (\eg, convLSTM \cite{shi2015convolutional, song2018pyramid}) to model the global-level inter-spectral characteristics of the spectral features. 
However, most modeling strategies typically employ 2D convolution, which primarily focuses on the overall temporal characteristics of the sequences, neglecting the continuity between adjacent sequences.
To address this, we propose a Neighborhood-wise Spectral Continuity Modeling (NeSCM) module.
It employs 3D convolution to recursively extract global continuity characteristics via memory units during the process of capturing temporal changes, effectively modeling the progressive variation patterns of spectral features.

As mentioned earlier, local correlation and global continuity characteristics describe the relationship between spectrum from different dimensions, and they are naturally complementary. 
Achieving adaptive fusion of these two features is crucial for more fully reconstructing the inter-spectral relationships.
Given the correspondence between these two features along the channel dimension, it is intuitive to adopt cross-attention mechanism to achieve image-level fusion, treating the channel as a sequence to model the inter-channel correlation.
However HSI reflects the inherent properties of materials, and different materials exhibit distinct spectral variation patterns. 
Therefore, the global progressive features should have different adaptive weights across various spatial regions during feature fusion, which is difficult to achieve with image-level fusion methods.
To address this issue, we improve the traditional image-level fusion strategy and design a Patch-wise Adaptive Fusion (PAF) module.
We preserve the ability to model inter-spectral sequence relationships while refining the spatial granularity during the fusion process. 
By dividing global and local features into patches along the spatial dimension, we can achieve patch-level adaptive fusion based on patch-wise similarity.

The main contributions can be summarized as follows.

\begin{itemize}
    \item The inter-spectral relationship in HSI is fully exploited from two dimensions, including the local correlation and the global continuity characteristics, so as to achieve efficient and accurate HSI reconstruction. Experimental analysis verifies that the proposed method achieves SOTA performance.
    \item We propose a grouped-mode GrSCM module to model the inter-spectral correlation within the local range, and design a temporal modeling-inspired NeSCM module to simulate the global continuity characteristic of the spectral features.
    
    \item We propose a regional-aware PAF module to fuse local correlation and global continuity features, which can collaboratively promote the network's ability to model the local and global inter-spectral relationships.
    
\end{itemize}

\section{Related Work} \label{sec2}

\subsection{Method Based on CNN}

In early image processing algorithms based on deep learning methods, CNN have often been employed to extract image features due to their exceptional ability to capture local patterns. Meanwhile, CNN have also demonstrated strong performance in HSI reconstruction tasks \cite{li2021survey}, and an increasing number of researchers are exploring the application of CNNs to address HSI reconstruction challenges \cite{alvarez2017adversarial, shi2018hscnn+, stiebel2018reconstructing, koundinya20182d, zhao2020hierarchical, li2021progressive, dian2023spectral}.
Alvarez-Gila \etal \cite{alvarez2017adversarial} pioneered the application of CNN for spectral reconstruction tasks, leveraging convolutional operations to extract latent contextual information within spatial local neighborhoods.
Koundinya \etal \cite{koundinya20182d} employs 2D convolution to extract spatial correlations in hyperspectral images, and utilizes 3D convolution to enhance the representation of spectral feature by exploiting inter-channel relationships. 
To enhance the representation of spectral features, Shi \etal \cite{shi2018hscnn+} employed residual blocks and dense connection layers for image feature extraction.
Zhao \etal \cite{zhao2020hierarchical} proposed a four-level hierarchical regression network to enlarge the receptive field of the network, enabling the model to incorporate contextual information during spectral feature reconstruction.

HSI data exhibits strong similarity across different spectral bands \cite{hang2021spectral}, prompting researchers to optimize network structures to exploit this characteristic.
The primary strategy involves employing channel attention to model the correlations between spectral bands \cite{li2022hasic, wu2023repcpsi, zhao2023hsgan, li2020hybrid, li2022drcr, zheng2021spectral, mei2023lightweight, li2021deep}.
To enhance the discriminative learning ability regarding band features, Li \etal \cite{li2022hasic} utilized channel attention and band attention modules to dynamically adjust feature responses at both the channel and band levels.
Wu \etal \cite{wu2023repcpsi} devised a lightweight coordinate-preserving proximity spectral-aware attention to capture spatial–spectral interdependences of intermediate features.
To enhance the generalization of the model across various RGB scenes, Zhao \etal \cite{zhao2023hsgan} introduced a channel attention mechanism in the discriminator.
Li \etal \cite{li2020hybrid} proposed 3D band attention to adaptively recalibrate bandwise feature responses for enhanced discriminative learning ability.
Li \etal \cite{li2022drcr} improved the dense connection module by integrating channel feature interdependencies with a re-calibration module.
Due to the similarity between adjacent spectral bands, some studies \cite{zheng2021spectral, mei2023lightweight} employ 1D convolution instead of multilayer perceptrons in channel attention, reducing computational complexity while modeling spectral similarities among neighboring bands.

While the channel attention module effectively models differences in responses across channels, it faces challenges in capturing more intricate correlation among spectral features, which are crucial for HSI reconstruction tasks.

\subsection{Method Based on Transformer}
The transformer \cite{vaswani2017attention}, owing to its advantage in modeling long-range dependencies, has been widely employed in research across various visual tasks \cite{dosovitskiy2020image, zamir2022restormer, yin2022csformer}.
Recently, researchers have also explored its application in HSI reconstruction tasks \cite{li2020adaptive, luan2024multi, cai2022mst++, du2023spectral, li2023mformer, wang2023learning}.
Considering that directly applying spatial self-attention operations on original features would introduce significant computational overhead, Li \etal \cite{li2020adaptive} designed a patch-level second-order non-local module, which divides the original image into sub-regions for attention computation.
To fully capture the similarity relationships between spectral bands in spectral features, Cai \etal ~proposed MST \cite{Cai_2022_CVPR} and MST++ \cite{cai2022mst++}.
These models partition tokens along the band dimension and establish similarity relationships between each spectral band feature and all others, enabling the network to effectively model spectral band similarities.
Du \etal \cite{du2023spectral} proposed a novel convolution and transformer joint network (CTJN). They were the first to integrate transformer with 3D convolution, thereby correlating local spatial information with global spectral information.
Li \etal \cite{li2023mformer} devised a dual spectral multi-head self-attention mechanism to enhance interaction between multiple heads and spectral dimensions.
To improve the specificity of spectral band modeling, Wang \etal \cite{wang2023learning} developed the Spectral-wise Re-Calibration module, applying channel attention alongside multi-head attention to enhance spectral feature representation.

Traditional transformer structures employ dense attention computations, which adopt a one-to-all approach to model complex spectral similarity correlations. 
This approach is relatively inefficient.
Moreover, relying solely on the transformer to model correlations between spectral band features, the network struggles to capture the progressive variation characteristics of spectral features along the global band dimension.

\begin{figure*}[!t]
        \centering
	\includegraphics[width=1\textwidth]{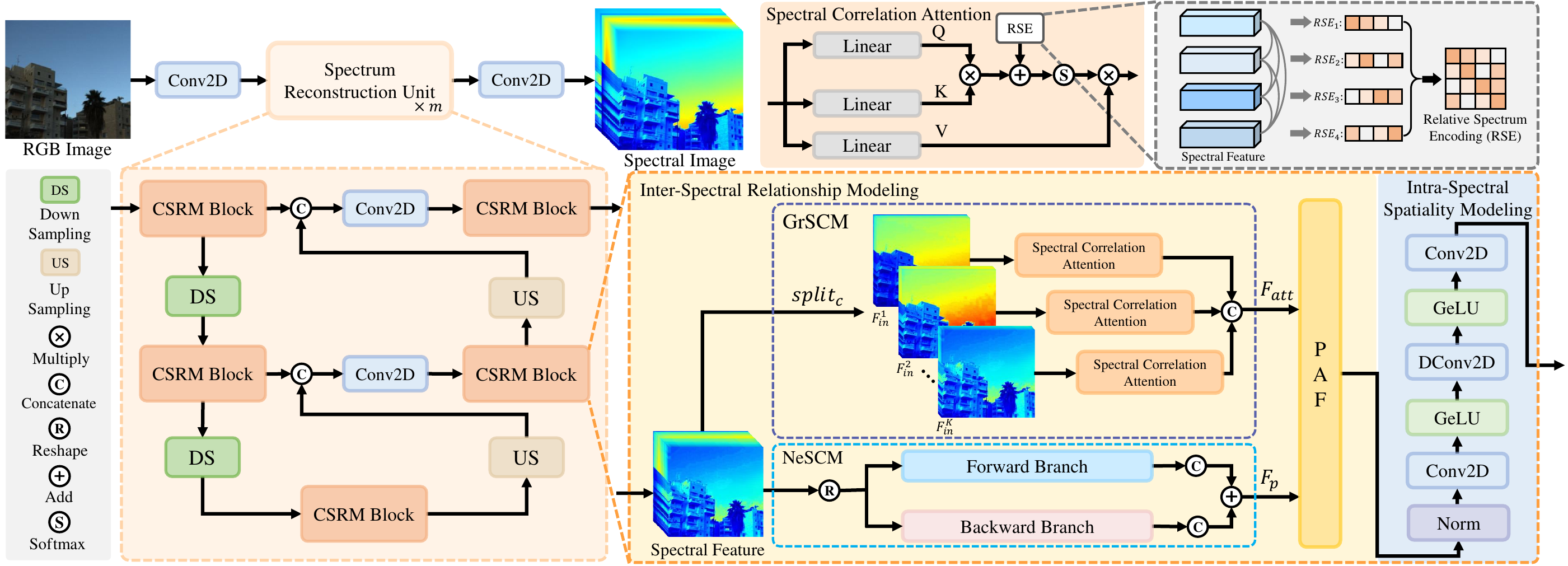}
    	\caption{The overall structure of the CCNet, which primarily consists of $m$ Spectrum Reconstruction Units based on the U-Net model \cite{ronneberger2015u}. 
    The Spectrum Reconstruction Unit consists mainly of multiple CSRM Blocks, with the core components of each CSRM block being Inter-Spectral Relationship Modeling module and Intra-Spectral Spatiality Modeling module.
    }
	\label{network}
\end{figure*}

\section{Methodology}\label{sec3}

The RGB camera is affected by the Spectral Response Function (SRF) while capturing images, and the captured images can be represented as:
\begin{equation}
    I_c(x,y)=\int_\lambda S^{\lambda}_{c}H_\lambda(x,y)
\end{equation}
where $\lambda$ denotes the bands within the visible light range, spanning from 400nm to 700nm, $c\in\{R,G,B\}$ represents the three channels corresponding to the RGB image and $S^\lambda_c\in \mathbb R^\lambda$ denotes the SRF of the channel $c$ for camera. 
$I_c(x,y)$ and $H_\lambda(x,y)$ respectively represent the visible light image with channel $c$ and the hyperspectral image with band $\lambda$. In practice, it can be as:
\begin{equation}
\label{eq:hsi2rgb}
    I_c(x,y)=\sum^B_{i=1} S^{\lambda}_{c}H_\lambda(x,y)
\end{equation}
where $B$ denotes the number of bands. 
The reconstruction of HSI from RGB images can be regarded as the inverse process of Eq. \ref{eq:hsi2rgb}, given $I_c$, to solve for $H$, and the specific process is also shown in Fig. \ref{process}.

\subsection{Network Architecture}\label{sec3.1}

The input to CCNet is RGB images, and the output is the reconstructed HSI. 
The core structure consists of $m$ serial spectral reconstruction units for spectral feature recovery, based on the U-Net model \cite{ronneberger2015u}, which converts the CNN into a Customized Spectral Relationship Modeling (CSRM) block capable of capturing intra- and inter-spectral relationships.
In the CSRM block, we first model the inter-spectral relationship. 
On the one hand, we design the Group-wise Spectral Correlation Modeling (GrSCM) module to model local spectral band similarity associations through group-wise attention. 
On the other hand, we develop the Neighborhood-wise Spectral Continuity Modeling (NeSCM) module to model the continuity of global spectral bands, which utilizes 3D convolution to recursively model progressive variation pattern at the global level. 
Finally, the Patch-wise Adaptive Fusion (PAF) module is used to fuse the features of the two modules mentioned above in order to jointly enhance the network's ability to model local correlation and global continuity features.
After modeling the inter-spectral relationship, we apply the Feed Forward Network (FFN) within the Transformer architecture to model intra-spectral spatiality, obtaining the final output of the CSRM block.

\subsection{Group-wise Spectral Correlation Modeling}

The Transformer structure has shown excellent performance in the HSI reconstruction task, largely due to its superior ability to model the similarity correlation between each band spectral feature and all other band features through the attention mechanism.
However, this process is inefficient.
HSI data, as shown in Fig. \ref{process}, reveals that the correlation between bands in the local region is high, while the correlation between more distant bands is relatively low.
This pattern is particularly evident in scenarios with a large number of spectral bands.
However, existing Transformer-based algorithms neglect this aspect, which will introduce some unnecessary correlations affecting the reconstruction effect, and will also introduce unnecessary computation.
To address this, as shown in Fig. \ref{network}, we design the GrSCM module, which divides the original spectral features along the channel dimension into several subregions and uses the group-wise attention mechanism within these subregions to model spectral similarity correlations. 
This enhances local spectral feature correlations while reducing unnecessary computational cost and optimization burden.

Assuming the size of the input spectral feature is $F_{in} \in \mathbb{R}^{H \times W \times C}$, we first perform a splitting operation along the channel dimension to obtain $F_{in}^i \in \mathbb{R}^{H \times W \times \frac{C}{k}}$, where $i \in \{1,2,\dots,k\}$ denotes the group index and $k$ is the number of groups.
Then, we sequentially perform intra-group linear mapping on the spectral features of each group:
\begin{equation}
    Q^i=F^i_{in}W^i_Q, K^i=F^i_{in}W^i_K, V^i=F^i_{in}W^i_V
\end{equation}
where $F^i_{in}\in \mathbb{R}^{(HW)\times \frac{C}{k}}$ represents the spectral features of the $i$-th group.
$W^i_Q$, $W^i_K$, and $W^i_V$ denote the parameters of the $i$-th linear mapping with dimensions of $\frac{C}{k}\times\frac{C}{k}$. 
The resulting $Q^i, K^i$ and $V^i$ are then used to calculate the inter-spectral attention weight for the $i$-th group.

For each group, we partition the spectral features into heads along the channel dimension, obtaining feature vectors $Q^i_h, K^i_h, V^i_h\in \mathbb{R}^{n\times (HW)\times \frac{C_{in}}{k}}$, where $C_{in}$ is the channel count for input features in the Spectrum Reconstruction Unit, and $n=\frac{C}{C_{in}}$ represents the number of heads. 
The head dimension can be interpreted as the feature dimension of the spectral bands. 
We compute the group-wise similarity between spectral band features to generate attention weights.
Considering the close relationship between the similarity of bands and their relative spectrum relationships, the similarity between band features can sometimes be reflected in their relative spectral positions.
Therefore, to better model the inherent correlation between the similarity of band features and their relative spectrum relationships, we introduce a learnable Relative Spectrum Encoding (RSE) when generating the attention weights to act as a relative position promopt.
Then, we apply softmax to obtain the normalized attention weights, followed by a weighted summation operation to derive the final attention features.
The specific process can be described as follows:
\begin{equation}
    F_{att_{h}}^i=\text{softmax}\left(\text{RSE}^i + \sigma \cdot \frac{{K^i_h}^TQ^i_h}{\lVert K^i_h \rVert\cdot\lVert Q^i_h \rVert}\right) \cdot V^i_h
\end{equation}
where $\text{RSE}^i\in \mathbb{R}^{n\times \frac{C_{in}}{k}\times\frac{C_{in}}{k}}$ is the relative spectrum encoding of the $i$-th group, respectively.
After that, we merge the head dimensions of the features $F_{att_{h}}^i$, obtaining the output features $F_{att}^i$ for each group.
These features are then merged along the group dimension to yield the output feature $F_{att}$.

This group-wise spectral modeling approach can not only strengthen the local correlation of spectral features, but also further reduce the computational cost and the optimization burden of the network.
For instance, assuming the input features have $C$ channels, the traditional MHA module strategy \cite{cai2022mst++, wang2023learning} requires calculating $C^2$ attention weights.
In comparison, the GrSCM module only requires calculating a total of $C^2/k$ attention weights, reducing the number of attention weights to be calculated and learned.

\begin{figure*}
        \centering
	\includegraphics[width=1.0\textwidth]{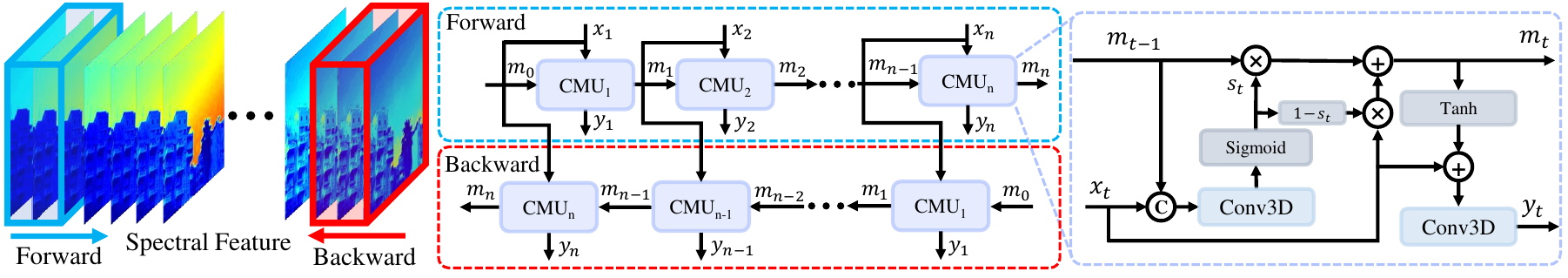}
	\caption{{The overall structure of NeSCM module, where the blue dashed lines and red dashed lines represent the forward branch and the backward branch, respectively. Each branch is composed of multiple CMUs.}}
	\label{SPM}
\end{figure*}

\subsection{Neighborhood-wise Spectral Continuity Modeling}

In addition to the local correlations, inter-spectral relationships also have the property of global continuity characteristics.
From the first spectral band to the last, the response values of HSI progressively change at each band rather than all at once.
However, the existing methods do not fully exploit and model this spectral attribute.
To this end, as shown in Fig. \ref{network}, we propose a Neighborhood-wise Spectral Continuity Modeling (NeSCM) module to effectively explore the continuity of the spectral features, mainly consisting of a forward branch and a backward branch. The specific structure of this module is shown in Fig. \ref{SPM}.

Following the GrSCM module's method of partitioning the head dimension, we divide the spectral features along the channel dimension, designating one dimension to represent the band dimension of the spectral features, resulting in features $F_{p_{in}} \in \mathbb{R}^{n\times H\times W\times C_{in}}$.
Inspired by the advantages of recursive modeling strategies (\eg, convLSTM \cite{shi2015convolutional, song2018pyramid}, convGRU \cite{wang2018learning}) for extracting global temporal characteristics from video data, we treat spectral features along the band dimension as a time series and adopt a recursive approach to model global temporal characteristics across spectral bands.
Most modeling strategies predominantly utilize 2D convolution to perform iterative analysis on individual sequence features, effectively modeling the overall temporal characteristics of the global sequence.
However, these approaches do not emphasize continuity between adjacent sequences, making it challenging to model the continuity of adjacent spectral features.
To address this, we improve the traditional recursive strategy by performing iterative analyses on multiple adjacent spectral bands simultaneously, and utilize 3D convolution to model the global-level progressive variation characteristics across spectral bands.

We select a window of size $s$ and perform a sliding window operation along the spectral dimension, extracting continuously distributed spectral segment features $\{x_1,x_2,\dots,x_{C_{in}}\}$, where $x_i\in \mathbb{R}^{n\times H\times W \times s}$. 
We then feed these features separately into forward branch and backward branch to extract global continuity features in two directions.
For forward branch, we initialize a memory feature $m_0$ and then pass each spectral segment feature along with the memory feature into the Continuous Memory Unit (CMU).
This process is performed recursively to update the memory feature and produce the output features for the current state.
For backward branch, the computational process is similar, except that the spectral segment features are fed in the reverse order compared to forward branch. 

For each CMU, we input the previous state memory feature $m_{t-1}$ and the current state spectral segment feature $x_t$, both having the same dimensions of $\mathbb{R}^{n\times H\times W\times s}$. 
To efficiently implement inter-spectral continuity modeling at the global level, we construct two gating strategies for each CMU, \ie, the forget gate and the output gate. 
The forget gate combines the historical memory feature $m_{t-1}$ and the current state spectral segment feature $x_t$ to generate forgetting weights, which discard some historical information from the memory features. 
Specifically, we concatenate $m_{t-1}$ and $x_t$ along the feature dimension, and apply 3D convolution with a $1\times1\times1$ kernel to generate the forgetting weights for the current unit:
\begin{equation}
    s_t = \text{Sigmoid}\left(\text{Conv3D}\left(\text{Concat}(m_{t-1}, x_t)\right)\right)
\end{equation}
where $s_t$ represents the forgetting weights of the current unit, and $\text{Conv3D}$ is the 3D convolution operation. Then, we update the memory sequence to obtain the memory feature $m_t$:
\begin{equation}
    m_t = s_t \times m_{t-1} + (1 - s_t) \times x_t
\end{equation}

Subsequently, for the output gate, we fuse the memory feature $m_t$ with the spectral segment feature $x_t$ and pass it through the 3D convolution with a $3\times3\times s$ kernel to model the spectral continuity characteristics:
\begin{equation}
    y_t = \text{Conv3D}\left(Tanh(m_t) + x_t\right)
\end{equation}
where $y_t$ represents output of the $t$-th CMU, with dimensions $\mathbb{R}^{n\times H\times W\times 1}$, which is referred to as progressive feature.
Both forward branch and backward branch yield $C_{in}$ progressive features. 
Then, these two sets of features are concatenated separately along the band dimension and then fused to generate the global progressive features $F_p$, which can represent the global-level progressive variation characteristics.
In this process, we utilize 3D convolution to simultaneously select multiple spectral segment features for iterative computation, enhancing the modeling capability of continuity between adjacent sequences while considering global temporal characteristics.
This enhances the network's ability to model global inter-spectral relationships.

\begin{figure*}
        \centering
	\includegraphics[width=1\textwidth]{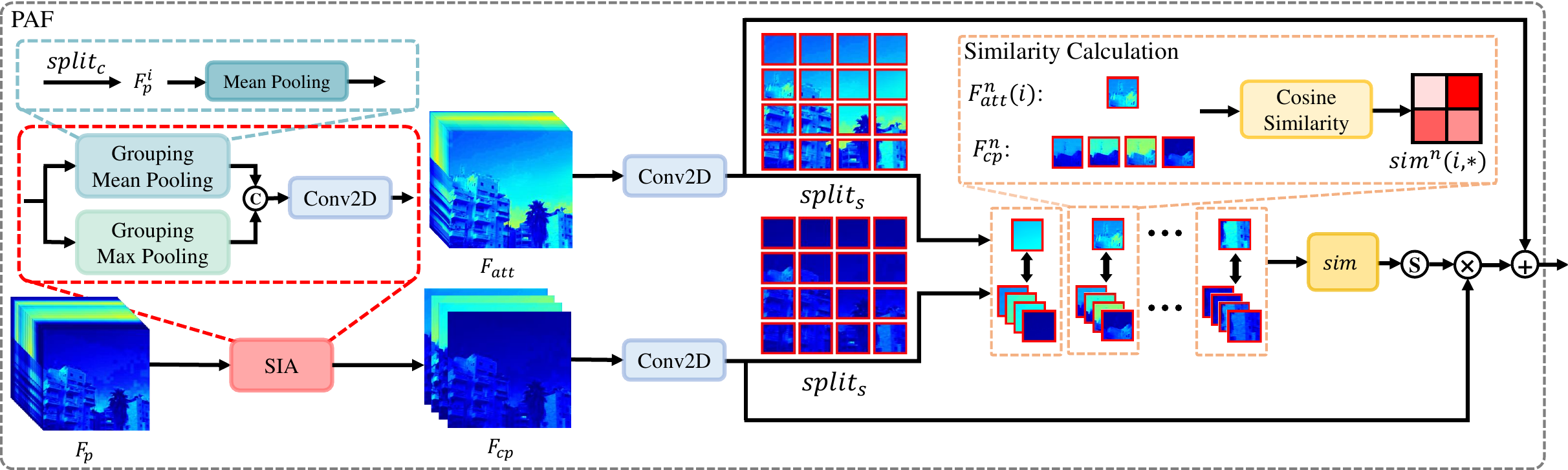}
	\caption{{The overall structure of PAF module. The similarity calculation in the orange rectangle box means that a $F_{att}^n(i)$ is calculated once with all $F_{cp}^n$, where $F_{cp}^n\in\{F_{cp}^n(1),F_{cp}^n(2),\dots, F_{cp}^n(k)\}$}}
	\label{SPFM}
\end{figure*}

\subsection{Patch-wise Adaptive Fusion}

The spectral features $F_{att}$ obtained from the GrSCM module efficiently construct spectral similarity correlations among band features within a local region.
Additionally, the global progressive features $F_p$ output by the NeSCM module capture the global-level progressive variation characteristics of the original spectral features.
These two sets of features emphasize different aspects and are complementary to each other.
It is crucial to model the inherent complementarity between these two features to generate comprehensive inter-spectral relationship features.
To this end, we design a Patch-wise Adaptive Fusion (PAF) module, which uses a patch-level fusion strategy to adaptively integrate $F_p$ into $F_{att}$, achieving adaptive feature fusion in local spatial regions and more comprehensively capturing the inter-spectral relationships.
The specific structure of the PAF module can be seen in Fig. \ref{SPFM}.

First, considering the group modeling strategy of $F_{att}$, we design a Spectrum Information Aggregation (SIA) module to aggregate spectral features using a group-based approach.
The grouping operation follows a rule similar to that of the GrSCM module, represented as $F_{p}^i=split_c(F_{p})$, where $F_{p}^i$ denotes the grouped progressive feature of the $i$-th group and $i\in\{1,2,\dots,k\}$ representing the group index.
Subsequently, within each group, both mean pooling and max pooling are applied along the channel dimension to the grouped progressive features, and a 2D CNN is used to fuse these two feature sets, generating the $i$-th compressed progressive feature $F_{cg}^i$.
This process can be represented as:
\begin{gather}
    F_{\text{mean}}^i = \frac{1}{C_p^i} \sum_{j=1}^{C_p^i} F_{p}^i(j) \\
    F_{\text{max}}^i = \max_j F_{p}^i(j) \\
    F_{cp}^i = \text{Conv2D}\left(\text{Concat}(F_{\text{mean}}^i, F_{\text{max}}^i)\right)
\end{gather}
where $C_p^i$ denotes the number of channels of the grouped progressive features for the $i$-th group.
Finally, we merge all $F_{cp}^i$ along the channel dimension to obtain the compressed progressive features $F_{cp}$, resulting in a size of $(H,W,k)$.

Then, we process $F_{att}$ and $F_{cp}$ separately through two 2D CNN operations with a kernel size of $1\times1$ for feature mapping.
Then, we divide the mapped features into patches along the spatial dimension, obtaining $F_{att}^n$ and $F_{cp}^n$, where $n$ represents the order of the patches.
Assuming the patch size is $r \times r$, we obtain a total of $\frac{HW}{r^2}$ patch features for a single channel's features.
Furthermore, by calculating the cosine similarity between patch features, we can obtain the attention weights:
\begin{gather}
    sim^n(i,j)=\frac{F^n_{att}(i)\cdot F^n_{cp}(j)}{\| F^n_{att}(i) \| \cdot \| F^n_{cp}(j) \|}\\
    w^n(i,*)=softmax(sim^n(i,*))
\end{gather}
where $sim^n(i,j)$ and $w^n(i,j)$ denote the similarity and attention weight, respectively, between the $i$-th feature $F^n_{att}(i)$ and the $j$-th compressed progressive features $F^n_{cp}(j)$. 
The $sim^n(i,*)$ indicates the similarity between the $i$-th channel's feature in $F_{att}^n$ and all compressed progressive features $F_{cp}^n$.

After obtaining the attention weights, we use them to integrate the $F_{cp}$ into the $F_{att}$:
\begin{equation}
    F_{out}=F_{att} + \underset{\forall{n,i}}{\text{Concat}}\left(\sum_{j=1}^{k}{F_{cp}^n(j) \cdot w^n(i,j)} \right)
\end{equation}
where $F_{out}$ represents the output feature of the module, and $\underset{\forall{n,i}}{\text{Concat}}$ denotes concatenating all patches sequentially along the channel dimension $i$ and spatial dimension $n$.
By dividing each band's spectral feature into patches along the spatial dimensions and performing feature fusion at the patch level, we can achieve adaptive feature fusion with spatial variations.
This fusion operation explores the intrinsic relationships between $F_{cp}$ and $F_{att}$, allowing the network to simultaneously model the local correlations and global progressive variation characteristics of the spectral features, thereby facilitating the generation of more comprehensive inter-spectral relationships.

\subsection{Loss Function}
Assuming $Y$ and $\hat{Y}$ represent the ground truth HSI and the reconstructed HSI, we employ two types of losses to optimize the overall network parameters, respectively.

(1) MRAE Loss: We utilize the MRAE loss to calculate the relative error between the reconstructed HSI and the ground truth HSI, optimizing the network parameters at a per-pixel granularity. The loss formula can be expressed as:
\begin{equation}
    L_{MRAE}(Y,\hat{Y})=\frac{1}{N}\sum_{i=1}^{N}{\frac{ \left| Y[i] - \hat{Y}[i] \right|}{Y[i]}}
\end{equation}
where $N$ indicates the total number of values in the HSI.

(2) Spectral Difference Loss: To further model the inter-spectral relationships in HSI, we propose a spectral difference loss. 
Specifically, we compute the difference between one band and all other bands in the reconstructed HSI using the $L_1$ distance, resulting in difference data of size $(C^2,H,W)$. Similarly, we compute the inter-band difference data for the ground truth HSI. The $L_1$ loss between these two sets of differences constitutes the spectral difference loss.
The specific loss formula can be expressed as:
\begin{gather}
    L_{Dif}(Y, \hat{Y}) =
    \frac{1}{C} \sum_{i=1}^{C}{\sum_{j\neq i}{\left| \lvert Y(i) - Y(j)\rvert - \lvert\hat{Y}(i) - \hat{Y}(j) \rvert \right|} }
\end{gather}

Finally, the total loss of the model can be expressed as:
\begin{equation}
    Loss=L_{MRAE} + \gamma L_{Dif}
\end{equation}
where $\gamma$ represents the weight of Spectral Difference Loss.

\begin{table*}[!h]
    \centering
    {
    \caption{Performance comparison with State Of The Arts over the NTIRE2022, NTIRE2020Real and NTIRE2020Clean datasets. Note that the best results are in bolded.}
    \begin{center}
        \renewcommand{\arraystretch}{1.3}
        \setlength{\tabcolsep}{2.5mm}{
            \begin{tabular}{c|ccc|ccc|ccc}
\hline
    \multirow{2}{*}{Method} & \multicolumn{3}{c|}{NTIRE2022} & \multicolumn{3}{c|}{NTIRE2020Real} & \multicolumn{3}{c}{NTIRE2020Clean} \\
    &  $\text{MRAE}\downarrow$  & $\text{RMSE}\downarrow$  & $\text{PSNR}\uparrow$ & $\text{MRAE}\downarrow$  & $\text{RMSE}\downarrow$  & $\text{PSNR}\uparrow$ & $\text{MRAE}\downarrow$  & $\text{RMSE}\downarrow$  & $\text{PSNR}\uparrow$ \\
  \hline
HSCNN+ \cite{shi2018hscnn+} & 0.3814 &  0.0588 &  26.36 & 0.0687 &  0.0182 &  35.89 & 0.0445 & 0.0170 & 37.53 \\
HRNet \cite{zhao2020hierarchical} &  0.3476 &  0.0550 & 26.89 &  0.0672 &  0.0178 & 36.01 & 0.0450 & 0.0161 & 37.75 \\ 
AWAN \cite{li2020adaptive}  &  0.2500 &  0.0367 & 31.22 &  0.0669 &  0.0174 & 36.35 & 0.0433 & 0.0151 & 38.46 \\ 
HDNet \cite{hu2022hdnet} & 0.2048 & 0.0317 & 32.13 & 0.0651 & 0.0165 & 36.65 & 0.0457 & 0.0159 & 38.13 \\ 
Restormer \cite{zamir2022restormer} & 0.1833 & 0.0274 & 33.40 & 0.0681 & 0.0163 & 36.54 & 0.0407 & 0.0134 & 39.57 \\
DRCR \cite{li2022drcr} & 0.1771 & 0.0259 & 33.82 & 0.0666 & 0.0171 & 36.31 & 0.0384 & 0.0121 & 39.62 \\ 
MST++ \cite{cai2022mst++}   & 0.1645 & 0.0248 & 34.32 & 0.0624 & 0.0157 & 37.05 & 0.0370 & \underline{0.0118} & 40.20 \\ 
RepCPSI \cite{wu2023repcpsi} & - & - & - & - & 0.0170 & 36.24 & - & 0.0123 & \underline{40.49}  \\ 
HySAT \cite{wang2023learning}  & \underline{0.1599} & \underline{0.0246} & \underline{34.47} & \underline{0.0589} & \underline{0.0142} & \underline{37.85} & \underline{0.0363} & 0.0129 & 40.35 \\
\hline
\textbf{CCNet(Ours)}    & \textbf{0.1556} & \textbf{0.0231} & \textbf{35.13} & \textbf{0.0580} & \textbf{0.0137} & \textbf{37.86} & \textbf{0.0323} & \textbf{0.0111} & \textbf{41.48} \\  
Percentage Gain & 2.69\% $\downarrow$ & 6.10\% $\downarrow$ & 1.91\% $\uparrow$ & 1.53\% $\downarrow$ & 3.52\% $\downarrow$ & 0.03\% $\uparrow$ & 11.02\% $\downarrow$ & 5.93\% $\downarrow$ & 2.45\% $\uparrow$ \\
\hline
\end{tabular}}
	\end{center}
	\label{table:coca}}
\end{table*}

\section{Experiments}\label{sec4}

\subsection{Datasets}

NTIRE2022 \cite{arad2022ntire} is a large-scale natural HSI dataset encompassing a diverse array of natural scene images. 
It comprises 1,000 HSIs, with 900 used for training, 50 for validation, and 50 for testing. 
Each HSI has a resolution of $482\times512$ pixels, with a spectral resolution of 10nm spanning 31 spectral bands ranging from 400nm to 700nm.
Following \cite{cai2022mst++, wang2023learning}, due to the unavailability of HSI label data for the test images, we employ the validation set here to evaluate the performance metrics of the model.

NTIRE2020 \cite{arad2020ntire} is a HSI reconstruction dataset proposed by the NTIRE2020 competition, which comprises two tracks: ``Clean" and ``Real". 
In the ``Clean" track, RGB images are in an uncompressed format, derived from ideal camera response parameters and HSI data conversion. 
On the other hand, images in the ``Real" track simulate real-world application scenarios, converted from actual uncalibrated camera response parameters and HSI data, while also incorporating simulated imaging noise. 
We use both NTIRE2020 Real and NTIRE2020 Clean datasets to evaluate the performance of our algorithm. 
Both datasets consist of 480 HSIs, with 450 used for training, 10 for validation, and 20 for testing, all maintaining the same spatial and spectral resolution as NTIRE2022.
We also use the validation data from these two datasets to verify the effectiveness of our algorithm.

Following the evaluation metrics in \cite{cai2022mst++}, we utilize three metrics to assess the HSI reconstruction capability of the model, including the Mean Relative Absolute Error (MRAE), the Root Mean Square Error (RMSE) and the Peak Signal-to-Noise Ratio (PSNR).
They can quantify the per-pixel discrepancy between the reconstructed HSI and the ground truth HSI.

\begin{figure*}[!h]
        \centering
	\includegraphics[width=1.0\textwidth]{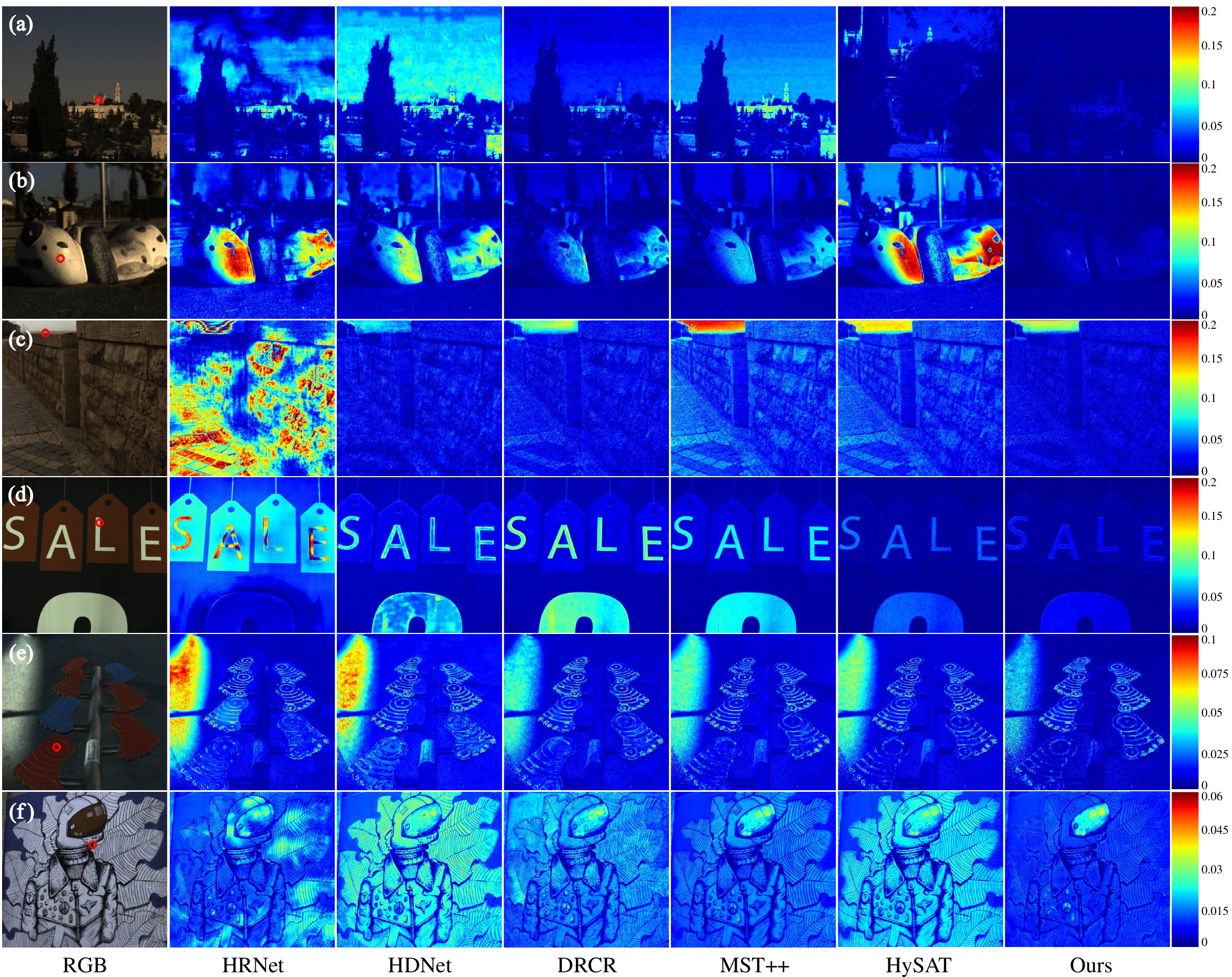}
	\caption{{The visualized $L_1$ reconstruction errors of different spectral reconstruction algorithms on different samples. The red area in the figure indicates a large error, and the blue area indicates a small error. For these samples, we randomly selected one point in each image, specifically marked by red circles in the RGB images, and visualized the spectral response curves, as shown in Fig. \ref{compare_line}.}}
	\label{compare_vis}
\end{figure*}

\subsection{Implementation Details} \label{Implementation}

For the hyperparameter settings, 
we set $k$ to 4 in the GrSCM module, meaning that a total of 4 groups are divided. Additionally, the number of initial spectral feature channels $C_{in}$ in the input of the Spectrum Reconstructed Unit is 32, and the number of these units $m$ is 3.
The initial patch size $r$ is set to 16 in the PAF module, and each time a feature undergoes downsampling, the size of $r$ is halved. 
The parameter $\gamma$ in the loss function is set to 0.1.
During the training phase, we normalize the pixel values of the RGB images to the range $[0,1]$ and apply random rotation and mirroring transformations for data augmentation. 
Following \cite{cai2022mst++}, we employ a multi-scale ensemble strategy to enhance the model's reconstruction capability. 
Our model is implemented in PyTorch \cite{paszke2017automatic}.
We use the Adam optimizer with $\beta_1=0.9$ and $\beta_2=0.999$. 
The initial learning rate is set to $4\times10^{-4}$, and we use a Cosine Annealing scheme to decay it to $10^{-5}$ over $200,000$ iterations on a single NVIDIA 4090Ti GPU.

\begin{figure*}[!h]
        \centering
	\includegraphics[width=1.0\textwidth]{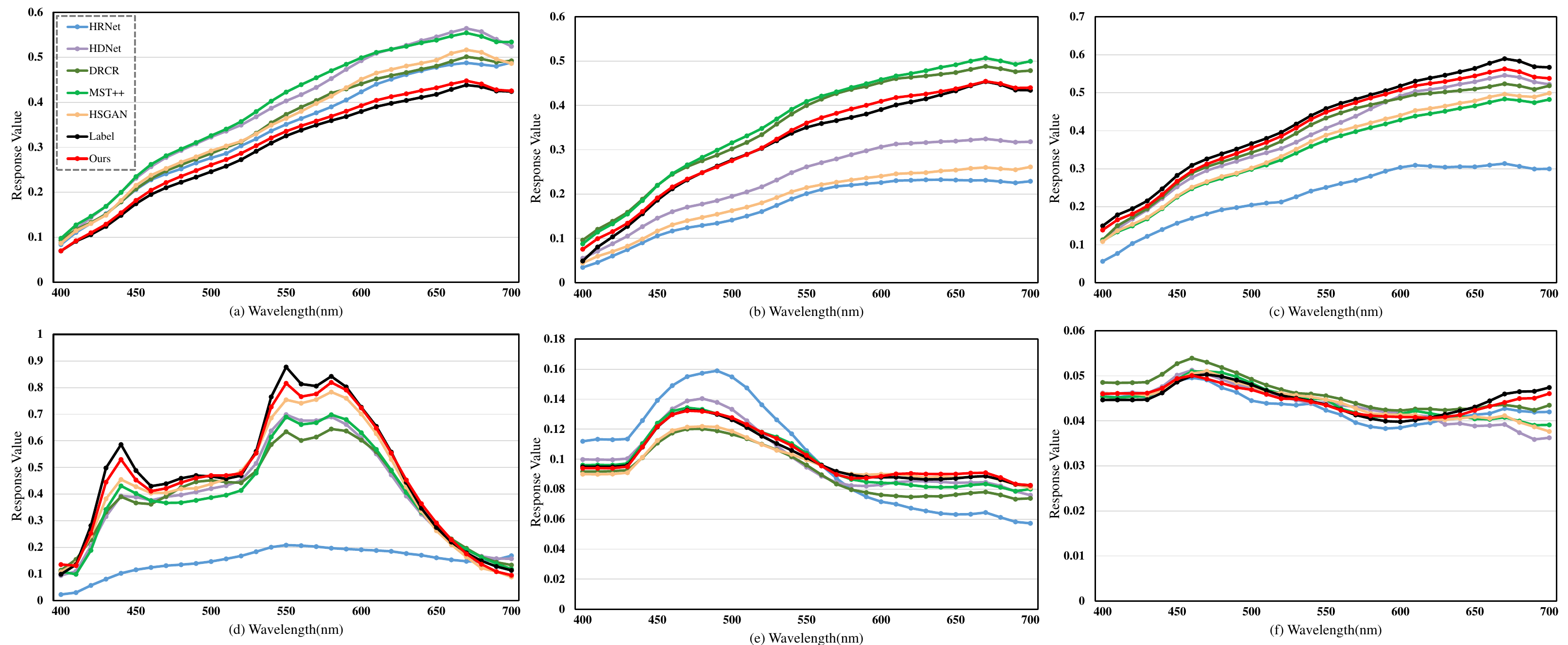}
	\caption{{The spectral response curves of different algorithms in a specific region of different samples correspond to the annotations in Fig. \ref{compare_vis}. The different colors of the curves indicate different reconstruction algorithms, as shown in the legend in the top left corner of subfigure (a). }}
	\label{compare_line}
\end{figure*}

\subsection{Comparison with the State-of-the-arts}
\subsubsection{Quantitative Analysis}

To demonstrate the effectiveness of the proposed algorithm, we compare it with several advanced HSI reconstruction algorithms, including HSCNN+ \cite{shi2018hscnn+}, HRNet \cite{zhao2020hierarchical}, AWAN \cite{li2020adaptive}, HDNet \cite{hu2022hdnet}, Restormer \cite{zamir2022restormer}, DRCR \cite{li2022drcr}, MST++ \cite{cai2022mst++}, RepCPSI \cite{wu2023repcpsi}, and HySAT \cite{wang2023learning}. 
MST++ and DRCR were the first and third place winners, respectively, in the Spectral Reconstruction from RGB track of the 2022 NTIRE competition, while RepCPSI and HySAT are advanced spectral reconstruction algorithms from 2023.
The specific comparison results are presented in Table \ref{table:coca}, where we have highlighted the best results in bold and underlined the suboptimal results. 
Based on the results from this table, our proposed method achieves optimal performance in HSI reconstruction.
Additionally, the percentage gain represents the incremental percentage of the optimal metric over the suboptimal metric
The proposed method demonstrates the most significant improvement on the NTIRE2020Clean dataset, achieving an 11.02\% reduction in MRAE and a 2.45\% increase in PSNR compared to the suboptimal results.
Moreover, the performance enhancement on the large-scale NTIRE2022 dataset is also notable, with the RMSE decreasing by 6.10\% relative to the best among other algorithms.

\subsubsection{Visualized Analysis}

To visually illustrate the effectiveness of the proposed algorithm, we compute the pixel-level L1 error between the reconstructed HSI and the ground truth HSI. 
These errors are averaged across all bands and visualized as heatmaps. 
Fig. \ref{compare_vis}(a)-(d) depict results from validation images in the NTIRE2022 dataset.
Fig. \ref{compare_vis}(e) represents validation images from the NTIRE2020Real dataset, while Fig. \ref{compare_vis}(f) represents validation images from the NTIRE2020Clean dataset.
In these heatmaps, a bluer color indicates smaller errors and higher quality reconstructed spectral images. 
The visual comparison demonstrates that the proposed algorithm outperforms existing algorithms in both overall reconstruction quality and detail.

Additionally, we select specific pixels marked by red circles in the above figures to plot spectral response curves for each algorithm. 
These curves are presented in Fig. \ref{compare_line}. 
From these plots, it is evident that the spectral response curves reconstructed by the proposed algorithm closely match the ground truth data.

\subsection{Ablation Study}

In order to analyze the contribution of each module, we conduct comprehensive ablation experiments on the NTIRE2022 dataset. 
These experiments validate the effectiveness of individual modules and analyze how hyperparameters influence the reconstruction accuracy of the algorithm.

\subsubsection{Modules Ablation Analysis}

In order to assess the effectiveness of each module, we conduct a module ablation analysis with detailed results presented in Tables \ref{table:ablation_gwa}-\ref{table:ablation_gpg}.

To validate the superiority of the proposed method in modeling inter-spectral relationships, we compared it with the MHA module. 
Here, we employed the same module to model intra-spectral spatial characteristics.
The experimental results are presented in Table \ref{table:ablation_gwa}, where FLOPS$_\text{s}$ represents the FLOPS of single inter-spectral relationships modeling module with an input size of $256 \times 256 \times 32$.
As shown in Table \ref{table:ablation_gwa}, when using the MHA module, the PSNR is $34.32 \text{dB}$, with a FLOPS of 0.44G for the single modeling module. 
In contrast, when employing the proposed method (including GrSCM, NeSCM, and PAF) to model inter-spectral relationships, there is a significant improvement in PSNR, with a slight decrease in FLOPS.
Additionally, the overall parameter of the proposed model is 1.61M, comparable to the MHA-based reconstruction algorithm.
This demonstrates that our designed module, compared to the traditional MHA module, effectively enhances HSI reconstruction quality without increasing computational complexity.

\begin{table}[!h]
\centering
\caption{The ablation experiment to verify the advanced of Inter-Spectral Relationships module.}
\label{table:ablation_gwa}
\renewcommand{\arraystretch}{1.2}
\begin{tabular}{ccccccc}%
\toprule
\makebox[0.02\textwidth][c]{MHA} & \makebox[0.04\textwidth][c]{Ours} & \makebox[0.02\textwidth][c]{RMAE} & \makebox[0.02\textwidth][c]{RMSE}& \makebox[0.02\textwidth][c]{PSNR}  & \makebox[0.055\textwidth][c]{FLOPS$_\text{s}$(G)}\\
\midrule
\checkmark &\quad & 0.1645 & 0.0248 & 34.32 & 0.44\\
\quad  & \checkmark & 0.1556 & 0.0231 & 35.13 & 0.42\\
\bottomrule %
\end{tabular}
\end{table}

To validate the effectiveness of the NeSCM module in modeling the global continuity of spectral features, we conducted a set of ablation experiment, as shown in Table \ref{table:ablation_bcr}, where ``Global Modeling'' refers to the modeling approach for global inter-spectral characteristics.
We retained the GrSCM module to model local correlations, and employed the PAF module to fuse local and global features when using the global strategy.
The ``b-convLSTM'' and ``b-convGRU'' are commonly used modules in video processing tasks for modeling global temporal characteristics \cite{hanson2018bidirectional, song2018pyramid, wang2018learning}. 
These modules perform iterative computations by sequentially inputting single-frame image data, enabling them to model global temporal information. 
By applying this method to HSI reconstruction, we sequentially input single spectral band features into the module for iterative computation, which allows for the modeling of global temporal characteristics. 
The PSNR of these two methods improved by 0.85 dB, demonstrating that modeling global temporal characteristics is essential for HSI reconstruction task.
However, traditional modules perform iterative operations on single spectral band features, making it difficult to model the continuity characteristics of spectral features.
We improved the iterative strategy by simultaneously inputting multiple adjacent spectral band features and using a 3D CNN to model global continuity characteristics.
This operation can produce global-level progressive variation features.
As demonstrated by the ``NeSCM" results in the table, this approach yields the highest HSI quality, with a PSNR reaching 35.13 dB. 
It can confirm that modeling global continuity is crucial for HSI reconstruction, and the NeSCM effectively models the inter-spectral continuity characteristics at the global level.

\begin{table}[!h]
\centering
\caption{The ablation experiment to verify the effectiveness of the NeSCM module in modeling the global continuity characteristics of spectral features.}
\label{table:ablation_bcr}
\renewcommand{\arraystretch}{1.2}
\begin{tabular}{cccc}%
\toprule
\makebox[0.12\textwidth][c]{Global Modeling} & \makebox[0.04\textwidth][c]{RMAE} & \makebox[0.04\textwidth][c]{RMSE}& \makebox[0.04\textwidth][c]{PSNR}\\
\midrule
\makebox[0.12\textwidth][c]{-} & 0.1887 & 0.0265 & 33.42 \\
\makebox[0.12\textwidth][c]{b-convLSTM}  & 0.1759 & 0.0245 & 34.27\\ 
\makebox[0.12\textwidth][c]{b-convGRU}  & 0.1701 & 0.0248 & 34.27 \\     
\makebox[0.12\textwidth][c]{NeSCM} & 0.1556 & 0.0231 & 35.13 \\
\bottomrule %
\end{tabular}
\end{table}

To demonstrate the effectiveness of the proposed PAF module in fusing $F_{att}$ and $F_p$, we conducted ablation experiments, with detailed results shown in Table \ref{table:ablation_gpg}. 
When using traditional fusion methods based on addition or concatenation \cite{chaib2017deep} to achieve feature fusion, the reconstruction accuracy of the HSI was relatively low, with PSNR values of only 33.72 and 34.11, respectively. 
This indicates that traditional fusion methods struggle to effectively capture the complementary relationship between the two sets of features.
For features with complementary characteristics, a popular approach is to use Cross Attention (CAttention) \cite{mohla2020fusatnet} to fuse multimodal features. 
We also conducted an experimental analysis using this method, achieving feature fusion along the channel dimension.
The PSNR reached 34.40 dB, significantly improving the reconstruction quality of HSI compared to traditional fusion strategies.
When we further refined this approach by replacing the image-level fusion with a patch-level fusion strategy, the PSNR results increased by an additional 0.52 dB. 
This demonstrates that the PAF enhances the effectiveness of feature fusion and improves the reconstruction quality of HSI by employing different adaptive weights for feature fusion in various regions.

\begin{table}[!h]
\centering
\caption{The ablation experiment to verify the effectiveness of the PAF module in fusing local and global inter-spectral features.}
\label{table:ablation_gpg}
\renewcommand{\arraystretch}{1.2}
\begin{tabular}{cccc}%
\toprule
\makebox[0.12\textwidth][c]{Fusion Method} & \makebox[0.03\textwidth][c]{RMAE} & \makebox[0.03\textwidth][c]{RMSE}& \makebox[0.03\textwidth][c]{PSNR} \\
\midrule
\makebox[0.06\textwidth][c]{Addition} & 0.1863 & 0.0260 & 33.72\\  
\makebox[0.06\textwidth][c]{Concatenation} & 0.1801 & 0.0263 & 34.11\\  
\makebox[0.06\textwidth][c]{CAttention} & 0.1698 & 0.0248 & 34.56\\ 
\makebox[0.06\textwidth][c]{PAF} & 0.1556 & 0.0231 & 35.13 \\
\bottomrule %
\end{tabular}
\end{table}

\subsubsection{Hyperparameter Sensitivity Analysis}

In the PAF, the patch size $r$ significantly influences the results of feature fusion. 
Both excessively large and small patch sizes hinder the integration of global progressive features with spectral features, as depicted in Table \ref{table:ablation_patch}. 
When the patch size is relatively large (\eg, 32 or 64), the spatial division of features becomes coarse. 
This makes it difficult to model spatial differences in spectral features across the channel dimension during fusion, leading to suboptimal reconstruction performance.
Conversely, when the patch size is small (\eg, 4 or 8), the spatial division becomes finer, enabling detailed modeling of spatial feature differences and improving feature diversity. 
However, this approach leads to a larger number of patches, significantly increasing the network's optimization burden. 
Consequently, achieving optimal performance becomes challenging, negatively affecting spectral reconstruction.
Setting the patch size to a moderate value (\eg, 16) allows the algorithm to strike a balance between modeling spatial differences and reducing the optimization burden. 
This approach results in relatively optimal spectral reconstruction performance.
\begin{table}[!h]
\centering

\caption{Comparison of performance results of different patch size on NTIRE2022.}
\label{table:ablation_patch}
\renewcommand{\arraystretch}{1.2}
\begin{tabular}{cccc}%
\toprule
\makebox[0.05\textwidth][c]{Size}  & \makebox[0.05\textwidth][c]{RMAE} & \makebox[0.05\textwidth][c]{RMSE}& \makebox[0.05\textwidth][c]{PSNR}       \\
\midrule
4 & 0.2396 & 0.0324 & 32.08   \\
8 & 0.1871 & 0.0267 & 33.70   \\ 
16 & 0.1556 & 0.0231 & 35.13  \\ 
32 & 0.1699 & 0.0237 & 34.72  \\ 
64 & 0.1704 & 0.0233 & 34.46  \\ 
\bottomrule %
\end{tabular}
\end{table}

\section{Conclusion}\label{sec5}
In this paper, we introduce the CCNet for reconstructing HSI from RGB images.
To model spectral correlations within a local range, we propose a grouped-mode GrSCM module to model the inter-spectral relationships locally. 
Additionally, we introduce a NeSCM module, inspired by temporal modeling, to recursively model the global continuity of spectral features.
Finally, we present a regional-aware PAF module that adaptively fuses the features output by the GrSCM and NeSCM modules.
We conduct extensive experimental analysis on three public HSI reconstruction datasets, demonstrating that the proposed algorithm achieves state-of-the-art performance.

{
\bibliographystyle{IEEEtran}
\bibliography{ref}
}

\vfill

\end{document}